\documentclass{article}
\usepackage{spconf,amsmath,graphicx}
\usepackage[utf8]{inputenc} 
\usepackage[T1]{fontenc}    
\usepackage{hyperref}       
\usepackage{url}            
\usepackage{booktabs}       
\usepackage{amsfonts}       
\usepackage{nicefrac}       
\usepackage{microtype}      
\usepackage{xcolor}         
\usepackage{soul}
\usepackage{graphicx}
\usepackage{amsmath}
\usepackage{algorithm}
\usepackage{algpseudocode}
\usepackage{floatpag}
\usepackage{enumitem}
\usepackage{arydshln}
\usepackage{svg}
\usepackage{pdfpages}
\usepackage{caption}
\usepackage{hyperref}
\usepackage{multirow}

\title{DIVA-VQA: Detecting Inter-frame Variations in UGC Video Quality}

\name{Xinyi Wang, Angeliki Katsenou, and David Bull\thanks{This work has been supported by the UKRI MyWorld Strength in Places Programme (SIPF00006/1).}}
\address{Visual Information Lab, School of Computer Science, University of Bristol, Bristol BS1 8UB, UK}

\begin{document}
\ninept
\maketitle
\begin{abstract}
The rapid growth of user-generated (video) content (UGC) has driven increased demand for research on no-reference (NR) perceptual video quality assessment (VQA).  NR-VQA is a key component for large-scale video quality monitoring in social media and streaming applications where a pristine reference is not available. This paper proposes a novel NR-VQA model based on spatio-temporal fragmentation driven by inter-frame variations. By leveraging these inter-frame differences, the model progressively analyses quality-sensitive regions at multiple levels: frames, patches, and fragmented frames. It integrates frames, fragmented residuals, and fragmented frames aligned with residuals to effectively capture global and local information. The model extracts both 2D and 3D features in order to characterize these spatio-temporal variations. Experiments conducted on five UGC datasets and against state-of-the-art models ranked our proposed method among the top 2 in terms of average rank correlation (DIVA-VQA-L: 0.898 and DIVA-VQA-B: 0.886). The improved performance is offered at a low runtime complexity, with DIVA-VQA-B ranked top and DIVA-VQA-L third on average compared to the fastest existing NR-VQA method. Code and models are publicly available at: \url{https://github.com/xinyiW915/DIVA-VQA}.
\end{abstract}
\begin{keywords}
NR-VQA, User-generated Content, Residual, SwinT, SlowFast
\end{keywords}
\section{Introduction}
\label{sec:intro}
Video Quality Assessment (VQA) is a critical component for optimizing user experience on sharing platforms such as YouTube, Instagram, and TikTok, which attract billions of daily video views~\cite{Omnicore23}. User-generated (video) content (UGC) is typically captured and encoded on consumer devices (smartphones or consumer-grade cameras) and is subsequently transcoded by streaming platforms to match a diverse range of devices and network conditions. These acquisition and transcoding processes result in UGC videos being delivered to users with degraded quality, which is manifested through artifacts such as blocking, ringing, and blurring~\cite{wang2019youtube}. 

Traditional full-reference (FR) VQA  methods rely on the availability of a pristine reference version for comparison with a distorted version. However, such reference content is generally not available for UGC videos. Hence, while FR methods, such as Video Multi-Method Assessment Fusion (VMAF)~\cite{li2016toward}, have achieved success in large-scale applications, they struggle to accurately evaluate the quality of transcoded content.  Distortions in UGC videos mainly arise from the limitations of recording devices, compression and transcoding, and transmission-induced artefacts~\cite{ying2021patch}. Specific content characteristics also play an important role in their perception by viewers~\cite{wang2009mean}.
Consequently, developing robust NR-VQA models that can accurately and consistently capture these types of artifacts and their perceptibility is urgently required.


Recent perceptual VQA models are typically based on machine learning approaches and follow a standard paradigm: extracting visual features related to perceived quality and predicting quality via regression or classification heads. Performance improvements have mainly been driven by enhanced feature extraction techniques, such as the integration of handcrafted features~\cite{tu2021ugc}, quality-aware pre-training~\cite{li2022blindly, wu2023exploring}, and advanced backbone networks~\cite{wu2023neighbourhood}, as well as on the design of quality score prediction heads. Although classical NR models~\cite{mittal2011blind, saad2014blind, korhonen2019two, tu2021ugc} perform well on small-scale datasets, they remain inadequate when addressing the complex spatio-temporal distortions associated with UGC video content. 

With the semantic awareness of deep neural networks (DNN) and the rise of large-scale UGC datasets, deep learning-based NR-VQA methods have become the mainstream approach. These models employ 2D-CNNs~\cite{li2019quality, tu2021rapique}, 3D-CNNs~\cite{liu2018end, sun2022deep}, and Transformers~\cite{wu2023discovqa, zhao2023zoom, wang2024relax} to analyze pixel variations caused by compression and correlate them with subjective quality scores. A critical aspect of video feature extraction is preserving both local details and global information effectively. However, DNN models with fixed input dimensions often require resizing videos, which can lead to the loss of local details. Cropping is commonly adopted ~\cite{ying2021patch} but tends to be overly localized for high-resolution videos, failing to represent overall quality. Distortions in UGC videos often exhibit transient characteristics~\cite{seshadrinathan2011temporal}, such as frame drops or focus shifts, which significantly impact perceived quality. Existing VQA models rely on networks pre-trained on large image classification datasets to extract frame-based or sampled frame features~\cite{li2019quality, you2021long, tu2021rapique, madhusudana2023conviqt}, demonstrating promising results. However, these methods introduce a distribution shift between the pre-trained tasks and the actual VQA prediction task, making it difficult to capture temporal distortions and adapt to the unique characteristics of UGC videos. Frame resizing further exacerbates quality degradation, prompting the development of VQA models that focus on selected fragments~\cite{wu2023neighbourhood, zhao2023zoom, wang2024relax, liu2024scaling}. 

The aim of NR-VQA is to emulate human perception of video quality, with influencing factors ranging from low-level details such as color and texture to high-level semantic content. To effectively model the complex factors associated with video quality, we propose a novel model DIVA-VQA that integrates three key feature components: raw frames, fragmented residuals, and fragmented frames aligned with the residuals. These components are designed to capture the spatio-temporal features of the video within different receptive fields. We design a dual-branch feature extractor to separately extract and combine the motion features and spatial features of video fragments, and employ a compact and efficient multilayer perceptron (MLP) regressor to perform model training and testing. 

We observed that the inter-frame variations between consecutive frames are generally similar, exhibiting considerable redundancy. Therefore, by analyzing the residual information between successive video frames, we can reveal the motion/inter-frame variations of moving regions and also reflect the spatial artifacts caused by compression in stationary regions. These residuals are highly correlated with perceptual changes in human vision and are capable of capturing key regions of interest. We further perform patch-level absolute difference ranking on the residuals, selecting the fragments with the greatest variation in the visual attention regions. This process has enabled us to develop an efficient quality-sensitive fragment extraction strategy, which establishes associations between frames and fragmented frames through residual patch position alignment, thereby leveraging these features to better capture the spatio-temporal quality variations between adjacent frames. Meanwhile, we combine the SlowFast\cite{feichtenhofer2019slowfast} and Swin Transformer (SwinT)\cite{liu2021swin} pre-trained models to extract 3D and 2D features of the video, respectively, thereby comprehensively capturing spatio-temporal features. Experimental results show that our method improves performance across extensive testing on five UGC datasets, while also ensuring the efficiency of the model.

The contributions of this framework are summarized as follows:
\begin{itemize}[leftmargin=*]
    \item We introduce a patch difference-based fragmentation strategy, exploring motion and inter-frame variations across consecutive video frames. This includes data preprocessing from frame to patch difference, and finally to fragmented frame.

    \item We propose a dual-branch feature extractor that captures spatio-temporal features from both the motion aspect and spatial information.     
    
    \item The proposed method has been extensively validated on public UGC datasets, demonstrating superior performance compared to other NR-VQA models both at accuracy and runtime complexity.
\end{itemize}

The remainder of this paper is structured as follows: Section~\ref{sec:method} outlines the proposed framework, while Section ~\ref{sec:exp} provides a detailed explanation of the experimental setup, configurations, and results. Finally, Section~\ref{sec:con} summarizes our findings and discusses potential directions for future research.


\section{Proposed Method}
\label{sec:method}

\begin{figure*}[htbp]
    \centering
    \includegraphics[width=.87\linewidth]{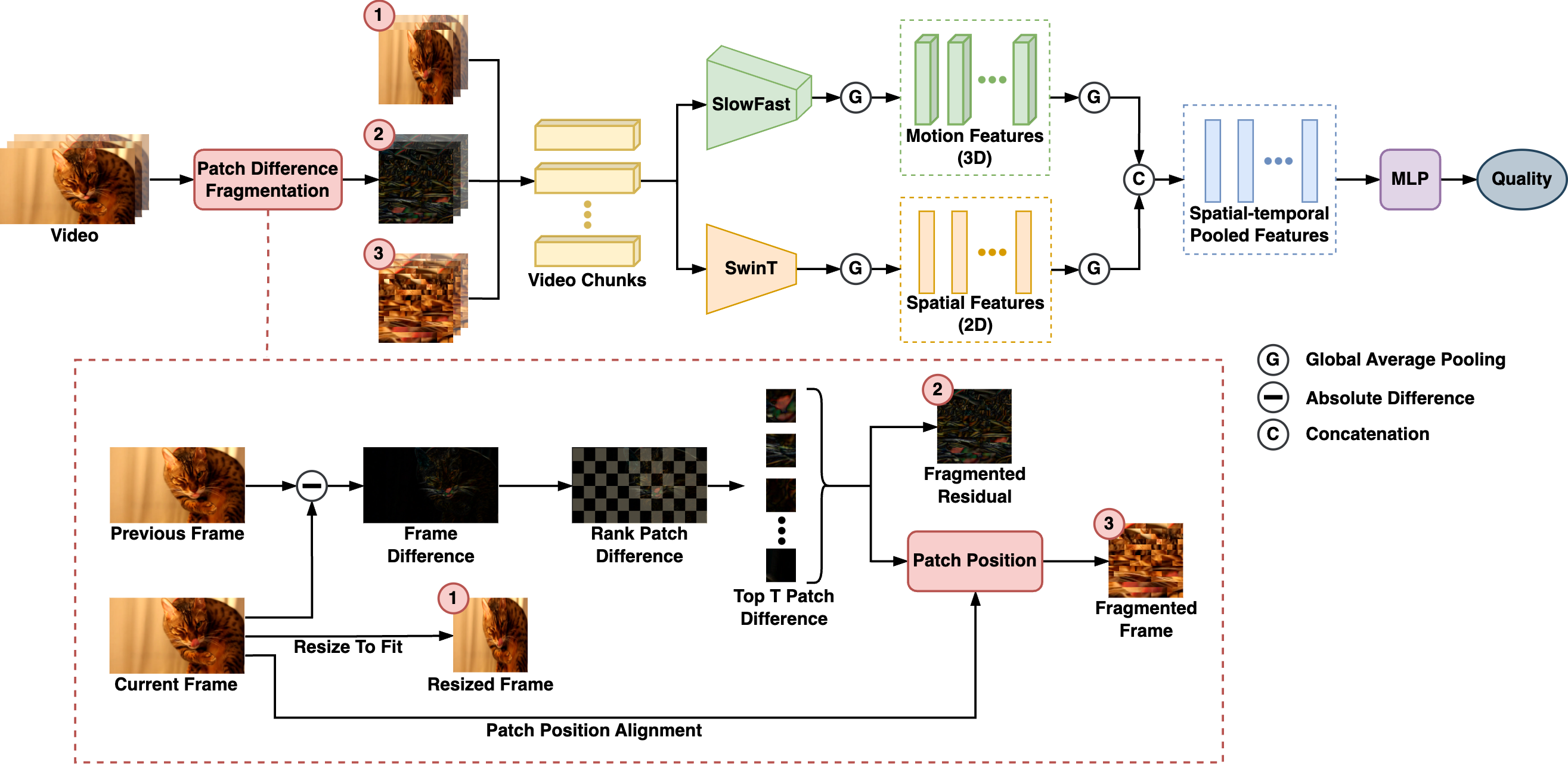}
    \caption{Overview of the proposed framework with two core modules: Patch Difference Fragmentation and Spatio-Temporal Feature Extraction. }
    \vspace{-2em}
  \label{fig: framework}
\end{figure*}

\subsection{Fragmentation using Patch Differences}
\label{ssec:frag-module}
We propose a patch difference fragmentation method inspired by ReLaX-VQA~\cite{wang2024relax}, to track motion changes in the video, and employ a patch alignment strategy to ensure the consistency of the positions between the extracted fragmented frames and fragmented residuals. This method divides the video at the frame level into patches and identifies significant inter-frame variations by calculating the differences within these regions, allowing for the extraction of key fragments that are most sensitive to distortion. 
Inter-frame patch variations localize the receptive field and can therefore better capture spatio-temporal visual attention at regions of interest—areas where meaningful motion or visual changes occur—rather than relying on a frame-level approach that applies attention globally, potentially diluting finer spatial details.

Given a video containing $N$ frames, after applying our fragmentation, three components are extracted for each frame: the resized video frame, a fragmented residual, and a fragmented frame aligned with its position. For current frame \( F_{\text{cur}}\) and previous frame \( F_{\text{pre}}\), we first calculate the frame difference \( R\) between the consecutive frames, which helps in eliminating redundant information:
\begin{equation}
\label{eqn:1}
R = | F_{\text{cur}} - F_{\text{pre}} |,
\end{equation}
where the residual \( R\) is subsequently used to extract key patches that are sensitive to video quality. Meanwhile, \( F_{\text{cur}}\) is adjusted to fit the target size \( s \times s\) required by the DNN model.

We divide \( R\) into multiple non-overlapping patches of size \( 
p \times p\). By quantifying the patch difference \( D_p\) for each patch, we rank all patches and select the top \( T\) patches with the highest magnitude or difference. These patches represent the candidate regions of significant changes within the frame. The difference measure \( D_p\) is defined as the sum of the absolute differences of all pixel values within the patch:
\begin{equation}
\label{eqn:2}
D_p = \sum_{x=1}^{p} \sum_{y=1}^{p} |P_{\text{cur}}(x, y) - P_{\text{pre}}(x, y)|,
\end{equation}
where \( P_{\text{cur}}(x, y)\) and \( P_{\text{pre}}(x, y)\) are the pixel values at position \( (x, y)\), respectively. The number of selected top \( T\) patches is derived from the patch size as the number of patches required to fit the target model's input size.
\begin{equation}
\label{eqn:3}
T = \left\lceil \frac{s^2}{p^2} \right\rceil \,.
\end{equation}

To ensure the consistency of the extracted spatio-temporal features, we introduce a patch position alignment strategy. For each patch in residuals, we retrieve the spatially aligned patch from frame \( F_{\text{cur}}\) using its coordinate information within the frame, forming a fragmented frame. These fragments capture the quality-sensitive regions of interest, avoiding attention to less error-prone areas, and reducing computational complexity.

\subsection{Spatio-Temporal Feature Extraction}
\label{ssec:feature-module}
To input video data into the pre-trained DNN models, we segment the video features into fixed-length chunks \( C_i\). For a video with \( N\) frames and a frame rate of \( f_r\), the number of divided chunks is \( M = \frac{N}{f_r}\). We define \( L_c \) as the length of each \( C_i\). Each \( C_i \) is a triplet comprising three components: 
\begin{equation}
\label{eqn:4}
C_i = \left( F_{\text{resized}}(C_i), R_{\text{frag}}(C_i), F_{\text{frag}}(C_i) \right),
\end{equation}
where \( F_{\text{resized}}(C_i) \) represents the resized frames, \( R_{\text{frag}}(C_i) \) represents the fragmented residual, and \( F_{\text{frag}}(C_i) \) represents the fragmented frame. The start and end indices of each \( C_i\) are defined as:
\(\text{start\_idx}_i = t \cdot f_r, \quad \text{end\_idx}_i = \text{start\_idx}_i + L_c,\)
where \( i \) is the chunk index, and \( t \) represents the temporal order of \( C_i\) in video sequence (e.g., \( t = 0, 1, 2, \dots \)). If the length of a chunk is less than \( L_c \), it is padded by repeating the last frame or fragment to ensure consistent length. The final processed set of video chunks is represented as:
\(
V_{\text{clips}} = \left[ C_1, C_2, \dots, C_M \right].
\)
To comprehensively capture rich spatio-temporal information, we employ a dual-branch feature extraction framework that combines the advantages of the SlowFast\cite{feichtenhofer2019slowfast} network and the SwinT\cite{liu2021swin}, enabling the extraction of quality-sensitive feature representations from videos.

The SlowFast model\cite{feichtenhofer2019slowfast} is a dual-pathway architecture. The ``Slow Pathway'' focuses on low-frequency features over longer temporal durations, capturing global motion changes, while the ``Fast Pathway'' targets high-frequency features over shorter time spans, capturing rapid and small changes. We extract slow and fast features and apply global average pooling to each. This dual-pathway design enables the model to capture multi-scale information in the temporal domain efficiently. Through this branch, we extract motion features from \( V_{\text{clips}}\):
{\footnotesize\begin{equation}
\label{eqn:5}
\begin{aligned}
\mathbf{Feats}_{\text{motion}} = \text{Concat}\bigl(&\text{GlobalAvgPool}\left(\text{SlowPath}(V_{\text{clips}})\right), \\
&\text{GlobalAvgPool}\left(\text{FastPath}(V_{\text{clips}})\right)\bigr).
\end{aligned}
\end{equation}
}

The SwinT\cite{liu2021swin} employs a hierarchical sliding window mechanism, which excels at extracting spatial features and long-range dependencies. We extract features using only the backbone network by removing its classification head. Similarly, we perform global pooling on the extracted feature maps to aggregate the local features. Through this branch, we extract spatial features from \( V_{\text{clips}}\):
\begin{equation}
\label{eqn:6}
\mathbf{Feats}_{\text{spatial}} = \text{GlobalAvgPool}\left(\text{SwinT}(V_{\text{clips}})\right).
\end{equation}
During the feature extraction stage, SlowFast and SwinT capture the temporal dynamics and spatial structure of the video from different dimensions, respectively. We further apply global average pooling to the extracted 3D and 2D features, followed by concatenation, resulting in unified spatio-temporal pooled features that simultaneously obtain both spatio-temporal details and global semantics.

\subsection{Quality Regressor}
\label{ssec:regressor-module}
We employed a compact and efficient multi-layer perceptron (MLP) regressor to fuse motion and spatial features, thereby enhancing the accuracy of video quality prediction. The model consists of three fully connected layers, with batch normalization, GELU activation, and dropout (0.1) incorporated between the layers to improve generalization and mitigate overfitting. The network maps the input features to 256-dimensional hidden units in the first layer, reduces the dimensionality to 128 in the second layer, and ultimately produces a single regression output through the final layer, yielding the objective quality
score. During training, we employed the stochastic gradient descent (SGD) optimizer with cosine annealing learning rate decay~\cite{loshchilov2016sgdr} and the stochastic weight averaging (SWA) technique~\cite{izmailov2018averaging} to ensure efficient optimization throughout the process.

We have adopted a composite loss function~\cite{wen2021strong}, which integrates Mean Absolute Error (MAE) and Rank Loss (MAERank Loss), combining two complementary components essential for optimizing both accuracy and ordinal consistency in VQA tasks. The first component is MAE, which minimizes the average absolute difference between the predicted values (\(y_{\text{pred}}\)) and the ground truth MOS (\(y_{\text{true}}\)), ensuring precise predictions:
\begin{equation}
\label{eqn:7}
    L_{\text{MAE}} = \frac{1}{N} \sum_{i=1}^N |y_{\text{pred}, i} - y_{\text{true}, i}| ,
\end{equation}
where \(y_i \in \mathbb{R}\) represents a quality score, \(N\) is the number of videos.

The second component, Rank Loss, ensures ordinal consistency by penalizing cases where the relative order of predictions does not match that of their ground truth values.
\begin{equation}
\label{eqn:8}
    L_{\text{Rank}} = \frac{1}{N^2} \sum_{i=1}^N \sum_{j=1}^N \max\left(0, \delta_{ij} - e(y_{\text{true}, i}, y_{\text{true}, j}) \cdot d_{ij}\right).
\end{equation}
where \(\delta_{ij} = |y_{\text{true}, i} - y_{\text{true}, j}|\) represents the absolute difference in ground truths, and \(d_{ij} = y_{\text{pred}, i} - y_{\text{pred}, j}\) represents the difference in predictions.
It calculates pairwise differences, weighted by a sign mask derived from the differences in the ground truth values. The function \(e(y_{\text{true}, i}, y_{\text{true}, j})\) is defined as follows:
\begin{equation}
\label{eqn:9}
    e(y_{\text{true}, i}, y_{\text{true}, j}) =
\begin{cases}
\,\,\,\, 1 & \text{, if } y_{\text{true}, i} \geq y_{\text{true}, j} \\
-1 & \text{, otherwise.}
\end{cases}
\end{equation}
An optional margin-based thresholding mechanism improves the robustness of Rank Loss by disregarding differences smaller than a predefined margin.
The composite loss \(L_{\text{MAERank}} \) assigns different weights: \(\text{mae}_w \) and \(\text{rank}_w\) to each loss to effectively balance precision and ranking objectives, thereby accelerating convergence.

\section{Experiments}
\label{sec:exp}
\begin{table*}[t]
    \centering
    \scriptsize
    \caption{Performance comparison of the evaluated NR-VQA models on the four NR-VQA datasets. The \textcolor{red}{\textbf{red}}, \textcolor{blue}{\textbf{blue}}, and \textbf{boldface} entries indicate the 1st, 2nd, and 3rd performance on each database for each performance metric, respectively.}
    \vspace{-1em}
    \label{tab: ComparisonToSoA}
    \begin{tabular}{@{}lllcccccccccc@{}}
    \toprule 
    \multicolumn{3}{c}{\textbf{Target Quality Dataset}} & \multicolumn{2}{c}{\textbf{CVD2014\cite{nuutinen2016cvd2014}}} & \multicolumn{2}{c}{\textbf{KoNViD-1k\cite{hosu2017konstanz}}} & \multicolumn{2}{c}{\textbf{LIVE-VQC\cite{sinno2018large}}} & \multicolumn{2}{c}{\textbf{YouTube-UGC\cite{wang2019youtube}}}  & \multicolumn{2}{c}{\textbf{Overall}}\\
    
    \cmidrule(lr){1-3} \cmidrule(lr){4-5} \cmidrule(lr){6-7} \cmidrule(lr){8-9} \cmidrule(lr){10-11} \cmidrule(lr){12-13}
    Type & Model & Pre-trained on & SRCC & PLCC & SRCC & PLCC & SRCC & PLCC & SRCC & PLCC & SRCC & PLCC\\
    \midrule
    \multirow{3}{*}{\textbf{Hand-crafted}} 
    & BRISQUE\cite{mittal2011blind} & NA &0.555 &0.553 &0.678 &0.675 &0.610 &0.665 &0.352 &0.377 &0.549&0.568\\
    & TLVQM\cite{korhonen2019two} & NA &0.540 &0.579 &0.762 &0.746 &0.813 &0.791 &0.680 &0.688 &0.699&0.701\\
    & VIDEVAL\cite{tu2021ugc} & NA  &0.766 &0.806 &0.807 &0.792 &0.773 &0.775 &0.781 &0.793 &0.782&0.792\\
    \hdashline
    \noalign{\vskip 2pt}
    \multirow{3}{*}{\textbf{Deep Learning}} 
    & VSFA\cite{li2019quality} & None &0.870 &0.868 &0.773 &0.775 &0.773 &0.795 &0.724 &0.743 &0.785&0.795\\
    & PVQ\cite{ying2021patch} & LSVQ\cite{ying2021patch} &- &- &0.791 &0.786 &0.827 &0.837 &- &- &0.809&0.812\\
    & BVQA\cite{li2022blindly} & None &0.872 &0.869 &0.834 &0.836 &0.834 &0.842 &0.818 &0.826 &0.840&0.843\\
    \hdashline
    \noalign{\vskip 2pt}
    \multirow{5}{*}{\textbf{Fragmentation}} 
    & FAST-VQA\cite{wu2023neighbourhood} & LSVQ\cite{ying2021patch} &0.891 &0.903 &0.891 &0.892 &\textbf{0.849} &0.862 &0.855 &0.852 &0.872&0.877\\
    & Zoom-VQA\cite{zhao2023zoom} & LSVQ\cite{ying2021patch} &- &- &0.877 &0.875 &0.814 &0.833 &- &- &0.846&0.854\\
    & DOVER\cite{wu2023exploring} & LSVQ\cite{ying2021patch} &- &- &\textcolor{red}{\textbf{0.909}} &\textcolor{blue}{\textbf{0.906}} &\textcolor{blue}{\textbf{0.860}} &0.875 &\textcolor{red}{\textbf{0.890}} &\textcolor{red}{\textbf{0.891}} &\textcolor{blue}{\textbf{0.886}}&\textbf{0.891}\\
    & ReLaX-VQA\cite{wang2024relax} & LSVQ\cite{ying2021patch} &\textbf{0.897} &\textcolor{red}{\textbf{0.929}} &0.872 &0.867 &0.847 &\textcolor{blue}{\textbf{0.888}} &0.847 &0.865 &0.866&0.887\\
    & SAMA\cite{liu2024scaling} & LSVQ\cite{ying2021patch} &- &- &\textbf{0.892} &0.892 &\textcolor{blue}{\textbf{0.860}} &\textbf{0.878} &\textcolor{blue}{\textbf{0.881}} &\textcolor{blue}{\textbf{0.880}} &\textbf{0.878}&0.883\\
    \hdashline
    \noalign{\vskip 2pt}
    \multirow{4}{*}{\textbf{Ours}} 
    & DIVA-VQA-B & None &0.869 &0.892 &0.856 &0.862 &0.825 &0.858 &0.821 &0.833 &0.843&0.861\\
    & DIVA-VQA-L & None &0.871 &0.896 &0.870 &0.873 &0.824 &0.860 &0.829 &0.835 &0.849&0.866\\
    & DIVA-VQA-B (wo/ fine-tune) & LSVQ\cite{ying2021patch} &0.840 &0.848 &0.849 &0.857 &0.807 &0.836 &0.734 &0.751 &0.807&0.823\\
    & DIVA-VQA-L (wo/ fine-tune) & LSVQ\cite{ying2021patch} &0.850 &0.854 &0.862 &0.866 &0.824 &0.849 &0.750 &0.765 &0.822&0.834\\
    & \textbf{DIVA-VQA-B (w/ fine-tune)} & LSVQ\cite{ying2021patch} &\textcolor{blue}{\textbf{0.900}} &\textcolor{blue}{\textbf{0.922}} &\textbf{0.892} &\textbf{0.900} &\textcolor{red}{\textbf{0.895}} &\textcolor{red}{\textbf{0.924}} &\textbf{0.858} &0.873 &\textcolor{blue}{\textbf{0.886}}&\textcolor{blue}{\textbf{0.905}}\\
    & \textbf{DIVA-VQA-L (w/ fine-tune)} & LSVQ\cite{ying2021patch} &\textcolor{red}{\textbf{0.911}} &\textbf{0.917} &\textcolor{blue}{\textbf{0.905}} &\textcolor{red}{\textbf{0.907}} &\textcolor{red}{\textbf{0.895}} &\textcolor{red}{\textbf{0.924}} &\textcolor{blue}{\textbf{0.881}} &\textbf{0.877} &\textcolor{red}{\textbf{0.898}}&\textcolor{red}{\textbf{0.906}}\\
    \bottomrule
    \end{tabular}
    \vspace{-1em}
\end{table*}

\subsection{Evaluation setup}
\label{ssec:setup}
\textbf{Training \& Benchmark Datasets:} We conducted intra-dataset performance evaluations on four state-of-the-art in-the-wild VQA datasets: CVD2014~\cite{nuutinen2016cvd2014}, KoNViD-1k~\cite{hosu2017konstanz}, LIVE-VQC~\cite{sinno2018large}, and YouTube-UGC~\cite{wang2019youtube}. Our model was built on the large-scale LSVQ dataset~\cite{ying2021patch}, comprising 38,793 videos, for feature extraction and training. The evaluation was conducted on the official test subsets of LSVQ (LSVQ\(_{\text{test}}\) and LSVQ\(_{\text{1080p}}\)). Furthermore, based on our LSVQ pre-trained model, we performed cross-dataset evaluations on the aforementioned NR-VQA benchmark datasets to assess the model's generalization capability. By fine-tuning the pre-trained model on these smaller-scale VQA datasets, the model successfully transferred and demonstrated substantial enhancements in performance. 

\noindent
\textbf{Evaluation Metrics:} We employed three metrics to evaluate the accuracy of quality predictions: Spearman Rank-Order Correlation Coefficient (SRCC), Pearson Linear Correlation Coefficient (PLCC), Kendall Rank-Order Correlation Coefficient (KRCC). These metrics assess the monotonicity, linearity, and overall accuracy of the predictions. To mitigate randomness, each experiment was repeated 21 times as in prior work, and the median result was reported~\cite{tu2021ugc}. SRCC and PLCC are highlighted in the performance comparison in Section \ref{ssec:comparison}, where higher values indicate better model performance.

\noindent
\textbf{Implementation Details:} We utilized SlowFast~\cite{feichtenhofer2019slowfast} pre-trained on Kinetics-400~\cite{kay2017kinetics} to extract motion features, and SwinT~\cite{liu2021swin} pre-trained on ImageNet-22k~\cite{krizhevsky2012imagenet} to extract spatial features. On the LSVQ dataset, we applied 10-fold cross-validation to enhance the model's generalization. While training on the LSVQ dataset, the model was run for 50 epochs using the SGD optimizer, with an initial learning rate of \(1 \times 10^{-1}\) and a weight decay of 0.005. When fine-tuning on smaller datasets, the model was trained for 200 epochs, with the learning rate of \(1 \times 10^{-2}\) and the weight decay of 0.0005. We utilized SWA with a learning rate consistent with the initial learning rate in the later stages of training. All datasets were split into training and testing sets with an 80\%-20\% ratio, and the batch size was set to 256. The loss function was configured with weight parameters \( \text{mae}_w = 0.6 \) and \( \text{rank}_w = 1.0 \). The best model was saved based on the minimum RMSE observed on the validation set. All experiments were conducted on a workstation equipped with an NVIDIA RTX 6000 Ada Generation GPU, a 28-core Intel(R) Core(TM) i7-14700K CPU, and 32 GB RAM. The proposed model was implemented in Python 3.10 using PyTorch.

\subsection{Performance comparison}
\label{ssec:comparison}
We evaluated the proposed model using the following four datasets:  CVD2014, KoNViD-1k, LIVE-VQC, and YouTube-UGC. Based on different pre-trained models and features, we designed two models:
\begin{itemize}[leftmargin=*]
    \item \textbf{DIVA-VQA Base (DIVA-VQA-B):} Utilizes the SlowFast R50-3D and Swin-Base pre-trained models with a patch size of 16. The feature dimension is 9984.

    \item \textbf{DIVA-VQA Large (DIVA-VQA-L):} Utilizes the SlowFast R50-3D and Swin-Large pre-trained models with a patch size of 16. The feature dimension is 11520.
\end{itemize}
 Results are reported in Table~\ref{tab: ComparisonToSoA} where we present the intra-dataset testing performance of the model across all datasets. Moreover, we provide the following performance comparison results, highlighting two key scenarios: (1) the cross-dataset testing performance of the model when trained on LSVQ features without fine-tuning (wo/ fine-tune), and (2) the intra-dataset performance improvements observed after fine-tuning the model to better fit the features of each dataset (w/ fine-tune). 

We employ the relevant metrics, SRCC and PLCC, to compare the current state-of-the-art (SOTA) NR-VQA models, which include hand-crafted statistical models, learning-based NR-VQA models, and the recently proposed fragmentation-based NR-VQA models. 

Our proposed model exhibits the best performance across all datasets, achieving the highest correlation scores on the CVD2014, KoNViD-1k, and LIVE-VQC datasets (SRCC or PLCC). Furthermore, all versions of our model outperformed traditional hand-crafted methods and frame-based deep learning approaches, further validating the effectiveness of the proposed fragmentation-based feature extraction method. Notably, our baseline version, DIVA-VQA-B, achieved SOTA performance on the LIVE-VQC dataset. Without fine-tuning, DIVA-VQA-B (wo/ fine-tune) and DIVA-VQA-L (wo/ fine-tune) models still achieved competitive results. For instance, on the KoNViD-1k dataset, DIVA-VQA-L (wo/ fine-tune) achieved a PLCC of 0.866, outperforming deep learning models specifically trained on the target dataset, with an improvement of $\Delta$PLCC = 0.03 compared to BVQA method. 

Compared to other NR-VQA methods that also employ fragmentation techniques, our model achieved the best performance (in terms of SRCC or PLCC) on the CVD2014, KoNViD-1k, and LIVE-VQC. Compared to the SOTA fragmentation model SAMA (an improved version of FAST-VQA), our model achieved improvements of $\Delta$SRCC = 0.035 and $\Delta$PLCC = 0.046 on the LIVE-VQC dataset. Compared to ReLaX-VQA, our model also achieved an improvement of $\Delta$PLCC = 0.04 on the KoNViD-1k dataset. Our model maintained excellent performance on multi-resolution datasets such as LIVE-VQC and YouTube-UGC, demonstrating the perceptual capability and strong generalization of our model across a wide range of video quality scenarios.

Additionally, in Table~\ref{tab: ComparisonLSVQ}, we present the evaluation results of our model on LSVQ. Our model demonstrated good performance on both the test subset and the 1080p high-resolution subset of LSVQ, achieving the highest PLCC on LSVQ\(_{\text{1080p}}\). 


\begin{table}[t]
    \centering
    \footnotesize
    \caption{Performance comparison when trained on LSVQ. \textcolor{red}{\textbf{Red}}, \textcolor{blue}{\textbf{blue}}, and \textbf{boldface} entries indicate 1st, 2nd and 3rd best, respectively.}
    \vspace{-1em}
    \label{tab: ComparisonLSVQ}
    \begin{tabular}{@{}lcccccccccccccc@{}}
    \toprule
    \multicolumn{2}{c}{\textbf{Testing Set}} & \multicolumn{2}{c}{\textbf{LSVQ\(_{\text{test}}\)}} & \multicolumn{2}{c}{\textbf{LSVQ\(_{\text{1080p}}\)}} \\
    \cmidrule(lr){1-2} \cmidrule(lr){3-4} \cmidrule(lr){5-6}
    Type & Model & SRCC & PLCC & SRCC & PLCC\\
    \midrule
    \multirow{3}{*}{\textbf{Hand-crafted}} & BRISQUE\cite{mittal2011blind} &0.579 & 0.576 &0.497 &0.531\\
    & TLVQM\cite{korhonen2019two} &0.772 &0.774 &0.589 &0.616\\
    & VIDEVAL\cite{tu2021ugc} &0.794 &0.783 &0.545 &0.554\\
    \hdashline
    \noalign{\vskip 2pt}
    \multirow{3}{*}{\textbf{Deep Learning}} & VSFA\cite{li2019quality} &0.801 &0.796 &0.675 &0.704\\
    & PVQ\cite{ying2021patch} & 0.827 &0.828 &0.711 &0.739\\
    & BVQA\cite{li2019quality} &0.852 & 0.855 &0.771 &0.782\\
    \hdashline
    \noalign{\vskip 2pt}
    \multirow{5}{*}{\textbf{Fragmentation}} & FAST-VQA\cite{wu2023neighbourhood} &0.876 &0.877 &0.779 &0.814\\
    & Zoom-VQA\cite{zhao2023zoom} &\textcolor{blue}{\textbf{0.886}} &0.879 &\textcolor{red}{\textbf{0.799}} &0.819\\
    & DOVER\cite{wu2023exploring} &\textcolor{red}{\textbf{0.888}} &\textcolor{red}{\textbf{0.889}} &\textcolor{blue}{\textbf{0.795}} &\textcolor{blue}{\textbf{0.830}}\\
    & ReLaX-VQA\cite{wang2024relax} &0.869 &0.869 &0.768 &0.810 \\
    & SAMA\cite{liu2024scaling} &\textbf{0.883} &\textcolor{blue}{\textbf{0.884}} &0.782 &\textbf{0.822}\\
    \hdashline
    \noalign{\vskip 2pt}
    \multirow{2}{*}{\textbf{Ours}} & \textbf{DIVA-VQA-B} &0.877 &0.877 &0.789 &\textcolor{red}{\textbf{0.832}}\\ 
    & \textbf{DIVA-VQA-L} &0.881 &\textbf{0.881} &\textbf{0.790} &\textcolor{blue}{\textbf{0.830}}\\
    \bottomrule
    \end{tabular}
\end{table}

\subsection{Ablation studies and complexity analysis}
\label{ssec:ablation}
We explored the impact of two key factors on performance: patch size and frame sampling rate.

\noindent
\textbf{Ablation study on patch size:} In Table~\ref{tab: Abla_patch}, we analyze how different input size adaptations for various pre-trained SwinT models influence performance. To this end, we compared the effect of fragments constructed with different patch sizes on the model's performance. The results show that using the same Swin-B pre-trained model, our fragment extraction method performs best when the patch size is 16 and the fragmented video frames are constructed at a size of 224×224. Compared to a patch size of 8, SRCC and PLCC improved by approximately 1.48\% and 1.83\%, respectively; compared to a patch size of 32, improved by approximately 0.63\% and 1.07\%, respectively. This result indicates that a moderate patch size can more effectively capture motion variations in features. Too small patches result in overly dense fragmented features, increasing interference from irrelevant redundant information. Too large patches lack fine-grained features, making it difficult to highlight the key role of spatio-temporal variations between frames in VQA. Additionally, the Swin-Large pre-trained model yields the best performance. However, increasing the input resolution (e.g., from 224×224 to 384×384) did not lead to an improvement in model performance.
\begin{table}[t]
    \centering
    \footnotesize
    \caption{Ablation study on \textit{patch size} and \textit{pre-train SwinT}.}
    \vspace{-1em}
    \label{tab: Abla_patch}
    \begin{tabular}{@{}lllcccc@{}}
    \toprule
    \multicolumn{3}{c}{\textbf{Testing Set}} & \multicolumn{4}{c}{\textbf{KoNViD-1k\cite{hosu2017konstanz}}} \\
    \cmidrule(lr){1-3} \cmidrule(lr){4-7}
    Patch size & Resolution & Pre-train Model & SRCC & KRCC & PLCC \\
    \midrule
    8 & 224 & Swin-Base &0.8438 &0.6568 &0.8466\\
    16 & 224 & Swin-Tiny &0.8370 &0.6485 &0.8447\\
    16 & 224 & Swin-Small &0.8462 &0.6655 &0.8577\\
    16 & 224 & Swin-Base &0.8563 &0.6723 &0.8621\\
    \textbf{16} & \textbf{224} & \textbf{Swin-Large} &\textcolor{red}{\textbf{0.8695}} &\textcolor{red}{\textbf{0.6840}} &\textcolor{red}{\textbf{0.8733}}\\
    32 & 224 & Swin-Base &0.8509 &0.6644 &0.8530\\
    \midrule
    \midrule
    8 & 384 & Swin-Base &0.8489 &0.6633 &0.8555\\
    16 & 384 & Swin-Base &0.8470 &0.6581 &0.8479\\
    16 & 384 & Swin-Large &0.8638 &0.6752 &0.8670\\
    32 & 384 & Swin-Base &0.8455 &0.6578 &0.8528\\
    \bottomrule
    \end{tabular}
\end{table}

\noindent
\textbf{Ablation study on frame sampling rate:} Table~\ref{tab: Abla_sam} shows that full-frame sampling achieves the best performance, with SRCC and PLCC improving by approximately 1.25\% and 0.71\%, compared to interval frame sampling. This result suggests that when integrating the spatio-temporal features of SlowFast and SwinT, full-frame sampling allows for a more comprehensive capture of motion and spatial information, thereby more accurately reflecting video quality. While interval sampling reduces computational load, it diminishes the ability to capture subtle motion information.
\begin{table}[t]
    \centering
    \footnotesize
    \caption{Ablation study on \textit{sub sampling}.}
    \vspace{-1em}
    \label{tab: Abla_sam}
    \begin{tabular}{@{}lllcccc@{}}
    \toprule
    &\multicolumn{2}{c}{\textbf{Testing Set}} & \multicolumn{4}{c}{\textbf{KoNViD-1k\cite{hosu2017konstanz}}} \\
    \cmidrule(lr){2-3} \cmidrule(lr){4-7}
    Sampling rate & Model & SRCC & KRCC & PLCC \\
    \midrule
    \textbf{all frames} & \textbf{DIVA-VQA-B} &\textcolor{red}{\textbf{0.8563}} &\textcolor{red}{\textbf{0.6723}} &\textcolor{red}{\textbf{0.8621}}\\
    every other frame  & DIVA-VQA-B &0.8457 &0.6612 &0.8560\\
    \bottomrule
        \end{tabular}
\end{table}

\begin{figure}[htbp]
    \centering
    \includegraphics[width=.9\linewidth]{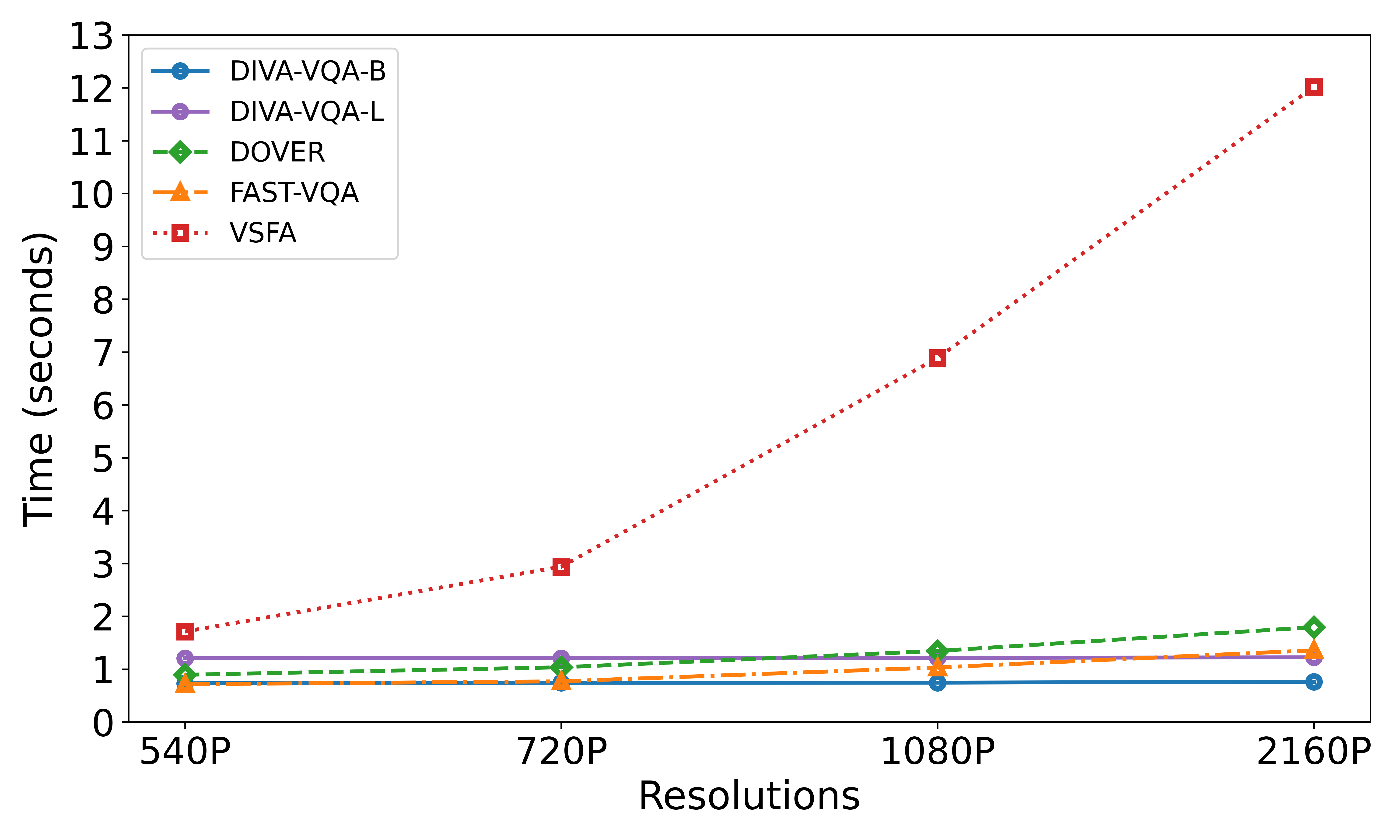}
    \caption{The comparison of running time (on GPU, averaged over ten runs) across different spatial resolutions.}
    \vspace{-2em}
  \label{fig: complexity}
\end{figure}

\noindent
\textbf{Runtime Complexity:} To ensure a fair comparison of the complexity of each method, all tests were executed on the same workstation (details in Section~\ref{ssec:setup}). We used the same video from the KoNViD-1k and conducted tests at different resolutions. Each test was repeated 10 times, with the average inference runtime of the model (in seconds) reported, as illustrated in Fig.~\ref{fig: complexity}, which shows the performance of the best NR-VQA models in terms of computation time (GPU). The results indicate that both DIVA-VQA-B and DIVA-VQA-L maintain stable runtimes across different resolutions, with minimal variation in runtime as the resolution increases. For example, DIVA-VQA-B takes 0.7320 seconds at 540P and only 0.7618 seconds at 2160P. Overall, DIVA-VQA-B achieves the fastest average runtime across all resolutions, while DIVA-VQA-L ranks third (Fast-VQA is second).


\section{Conclusion and Future Work}
\label{sec:con}
In this paper, we propose a novel NR-VQA model that addresses the challenges posed by complex spatiotemporal distortions encountered in UGC videos. 
Our method introduces ranked patch differences between consecutive video frames to generate fragmented video chunks, effectively improving the alignment between video frames and fragments, thereby strengthening the model's ability to perceive complex spatio-temporal distortions. 
Additionally, we employed a compact MLP regressor to train and evaluate the extracted video features. We 
trained our model on the large-scale LSVQ dataset, achieving consistently high performance on its test subsets. Furthermore, we fine-tuned the pre-trained model based on LSVQ video features and validated its performance on four publicly available UGC video datasets. Performance comparisons demonstrate that our method outperforms existing NR-VQA methods on average, achieving an SRCC of 0.898 and a PLCC of 0.906 across all datasets. Notably, both models come with low runtime complexity with DIVA-VQA-B ranking as the fastest across all resolutions.
Future research will focus on integrating textual descriptions of the videos that will add semantic context to the perceptual quality assessment.


\vfill\pagebreak
\bibliographystyle{IEEEbib}
\bibliography{strings,refs}

\end{document}